\documentclass[letterpaper,twocolumn,aps,prl,groupedaddress,superscriptaddress]{revtex4}
\usepackage[utf8]{inputenc}
\usepackage{color}
\usepackage{amstext}
\usepackage{graphicx}

\makeatletter

\@ifundefined{textcolor}{}
{%
 \definecolor{BLACK}{gray}{0}
 \definecolor{WHITE}{gray}{1}
 \definecolor{RED}{rgb}{1,0,0}
 \definecolor{GREEN}{rgb}{0,1,0}
 \definecolor{BLUE}{rgb}{0,0,1}
 \definecolor{CYAN}{cmyk}{1,0,0,0}
 \definecolor{MAGENTA}{cmyk}{0,1,0,0}
 \definecolor{YELLOW}{cmyk}{0,0,1,0}
}

\usepackage{dcolumn}
\usepackage{bm}

\usepackage{xcolor}

\makeatother

\begin{document}

\title{Neutron-scattering measurements of the spin excitations in LaFeAsO and 
Ba(Fe$_{0.953}$Co$_{0.047}$)$_{2}$As$_{2}$: Evidence for a sharp enhancement of 
spin fluctuations by nematic order}

\author{Qiang Zhang}

\affiliation{Ames Laboratory, Ames, IA, 50011, USA}

\affiliation{Department of Physics and Astronomy, Iowa State University, Ames,
IA, 50011, USA}

\author{Rafael M. Fernandes}

\affiliation{School of Physics and Astronomy, University of Minnesota, Minneapolis,
MN 55455, USA}

\author{Jagat Lamsal}

\affiliation{Ames Laboratory, Ames, IA, 50011, USA}

\affiliation{Department of Physics and Astronomy, Iowa State University, Ames,
IA, 50011, USA}

\author{Jiaqiang Yan}

\affiliation{Oak Ridge National Laboratory, Oak Ridge, Tennessee 37831, USA}

\author{Songxue Chi}

\affiliation{Oak Ridge National Laboratory, Oak Ridge, Tennessee 37831, USA}

\author{Gregory S. Tucker}

\affiliation{Ames Laboratory, Ames, IA, 50011, USA}

\affiliation{Department of Physics and Astronomy, Iowa State University, Ames,
IA, 50011, USA}

\author{Daniel K. Pratt}

\affiliation{NIST Center for Neutron Research, National Institute of Standards
and Technology, Gaithersburg, Maryland 20899-6102, USA}

\author{Jeffrey W. Lynn}

\affiliation{NIST Center for Neutron Research, National Institute of Standards
and Technology, Gaithersburg, Maryland 20899-6102, USA}

\author{R. W. McCallum}

\affiliation{Ames Laboratory, Ames, IA, 50011, USA}

\affiliation{Department of Materials Sciences and Engineering, Iowa State University,
Ames, Iowa 50011, USA}

\author{Paul C. Canfield}

\affiliation{Ames Laboratory, Ames, IA, 50011, USA}

\affiliation{Department of Physics and Astronomy, Iowa State University, Ames,
IA, 50011, USA}

\author{Thomas A. Lograsso}

\affiliation{Ames Laboratory, Ames, IA, 50011, USA}

\affiliation{Department of Materials Sciences and Engineering, Iowa State University,
Ames, Iowa 50011, USA}

\author{Alan I. Goldman}

\affiliation{Ames Laboratory, Ames, IA, 50011, USA}

\affiliation{Department of Physics and Astronomy, Iowa State University, Ames,
IA, 50011, USA}

\author{David Vaknin}

\affiliation{Ames Laboratory, Ames, IA, 50011, USA}

\affiliation{Department of Physics and Astronomy, Iowa State University, Ames,
IA, 50011, USA}

\author{Robert J. McQueeney}

\email{mcqueeneyrj@ornl.gov}

\affiliation{Ames Laboratory, Ames, IA, 50011, USA}

\affiliation{Department of Physics and Astronomy, Iowa State University, Ames,
IA, 50011, USA}

\affiliation{Oak Ridge National Laboratory, Oak Ridge, Tennessee 37831, USA}

\date{\today}
\begin{abstract}
Inelastic neutron scattering was employed to investigate the impact
of electronic nematic order on the magnetic spectra of LaFeAsO and
Ba(Fe$_{0.953}$Co$_{0.047}$)$_{2}$As$_{2}$. These materials are
ideal to study the paramagnetic-nematic state, since the nematic order,
signaled by the tetragonal-to-orthorhombic transition at $T_{{\rm S}}$,
sets in well above the stripe antiferromagnetic ordering at $T_{{\rm N}}$.
We find that the temperature-dependent dynamic susceptibility displays
an anomaly at $T_{{\rm S}}$ followed by a sharp enhancement in the
spin-spin correlation length, revealing a strong feedback effect of
nematic order on the low-energy magnetic spectrum. Our findings can
be consistently described by a model that attributes the structural/nematic
transition to magnetic fluctuations, and unveils the key role played
by nematic order in promoting the long-range stripe antiferromagnetic
order in iron pnictides. 
\end{abstract}


\maketitle
One of the most interesting features of the ``122'' (e.g. BaFe$_{2}$As$_{2}$)
and ``1111'' (e.g. LaFeAsO) families of iron-based superconductors
is the intimate coupling between superconductivity (SC), stripe antiferromagnetic
order (AFM), and the tetragonal-to-orthorhombic structural transition
\cite{review1,review2,review3,review4,review5}. For example, in both
families, chemical substitutions on the transition metal site, such
as Co and Ni, suppress the AFM ordering and the structural transition
and, over a limited range of doping, promote SC \cite{review4}. For
underdoped BaFe$_{2}$As$_{2}$, evidence of a direct competition
between AFM and SC has been presented \cite{Pratt2009,Christianson09,Fernandes_Pratt2010,Avci11,Dai12}
in addition to a suppression of the orthorhombic distortion below
the superconducting transition temperature $T_{{\rm C}}$ \cite{Nandi10,Bohmer12}.
Despite this competition between SC and long-range magnetic/orthorhombic
order, SC generally arises when large AFM/structural fluctuations
are present \cite{reviews_pairing}, a feature that attests the intricate
relationship between these three intertwined phases \cite{Fradkin14}.

While these previous studies have focused on the impact of SC on the
magnetic and orthorhombic phases, the interplay between these two
ordered states has been a topic of intense debate \cite{Fernandes2014}.
For the parent compounds of the ``122'' family, the magnetic transition
temperature ($T_{{\rm N}}$) practically coincides with the structural
distortion at $T_{{\rm S}}$ \cite{Li2009,Kim11,Birgeneau11}, whereas
in the Co-underdoped BaFe$_{2}$As$_{2}$ and in the parent compounds
of the ``1111'' family, such as LaFeAsO, the orthorhombic distortion
occurs well above $T_{{\rm N}}$ \cite{Li2011,Zhang}. The structural
transition has been proposed to be driven by electronic correlations
\cite{Fisher12} -- associated with either spin 
\cite{Kivelson,Xu2008,Fernandes2010,Fernandes2012b,Dagotto13}
or charge/orbital degrees of freedom \cite{w_ku10,Devereaux10,Phillips12,Kontani12}
-- giving rise to the so-called nematic phase in the temperature range
between $T_{{\rm S}}$ and $T_{{\rm N}}$. This electronic nematic
phase is characterized not only by a weak in-plane structural anisotropy
manifested by distinct $a$ and $b$ lattice constants \cite{Nandi10},
but also by large in-plane anisotropies in many electronic properties,
such as resistivity \cite{Chu2010,Tanatar10,Fisher12}, optical conductivity
\cite{Dusza2011,Nakajima2011,Jigang14}, thermopower \cite{Jiang2013},
uniform susceptibility \cite{Matsuda12,Cao14}, and charge correlations
\cite{Li2011,Zhang2013,Kim13}. Previous ARPES \cite{ZXshen11,DLFeng12,Yi2012,pseudogap_Matsuda},
STM \cite{Davis10,Rosenthal13}, and Raman \cite{Gallais13} studies
focused on how nematic order affects the normal-state electronic spectrum
and, in particular, the charge and orbital degrees of freedom. However,
little is known about how nematic order affects the low-energy magnetic
fluctuations\cite{Ma_NaFeAs,Imai12,PDai1,PDai2,PDai3}, which are
particularly important for the formation of the SC state \cite{reviews_pairing}.

Here we perform inelastic neutron scattering (INS) experiments to
elucidate the evolution of the magnetic spectrum across the nematic
transition in single crystals of LaFeAsO and Ba(Fe$_{0.953}$Co$_{0.047}$)$_{2}$As$_{2}$,
focusing on the behaviors of the imaginary part of the dynamic magnetic
susceptibility $\chi''(\textbf{Q},E)$ and of the spin-spin correlation
length $\xi$ as a function of temperature. These two systems exhibit
an orthorhombic distortion whose onset is well separated from the
stripe AFM ordering\cite{Pratt2009,Yan2009,Li2011,Ni2008}, enabling
the survival of the nematic phase over a considerable temperature
range. Our measurements in twinned samples find clear anomalies in
the magnetic spectrum at $T_{{\rm S}}$. In particular, we find that
not only is the overall low-energy magnetic intensity enhanced below
$T_{{\rm S}}$, but also that the spin-spin correlation length undergoes
a sharp increase at the nematic transition temperature, in contrast
with what one expects from a typical AFM system. This effect 
reveals
a cooperative relationship between nematicity and magnetism, in agreement
with theoretical predictions from models that attribute the nematic
transition to a spontaneous symmetry breaking driven by magnetic fluctuations
\cite{Kivelson,Xu2008,Fernandes2010,Fernandes2012b,Dagotto13}.

The LaFeAsO and Ba(Fe$_{0.953}$Co$_{0.047}$)$_{2}$As$_{2}$ crystals
were grown using a flux technique as previously described\cite{Yan2009,Ni2008}.
Dozens of small single-crystals of LaFeAsO with a total mass of approximately
600 mg were co-aligned in  the (\textit{H}0\textit{L}) plane within
$\sim2$ degrees mosaicity. Hereafter, unless otherwise noted with
a subscript ``T'', we use orthorhombic notation. A large single
crystal of Ba(Fe$_{0.953}$Co$_{0.047}$)$_{2}$As$_{2}$ with a mass
of $\approx700$ mg was also aligned in the (\textit{H}0\textit{L}) 
plane for
the investigation. The elastic and inelastic neutron measurements
on LaFeAsO and Ba(Fe$_{0.953}$Co$_{0.047}$)$_{2}$As$_{2}$ were
performed on the HB3 spectrometer (located at the High Flux Isotope
Reactor at Oak Ridge National Laboratory) and BT-7 triple-axis neutron
spectrometer at the NIST Center for Neutron Research {\cite{Lynn2012}},
respectively.

\begin{figure}
\centering \includegraphics[width=2.5in]{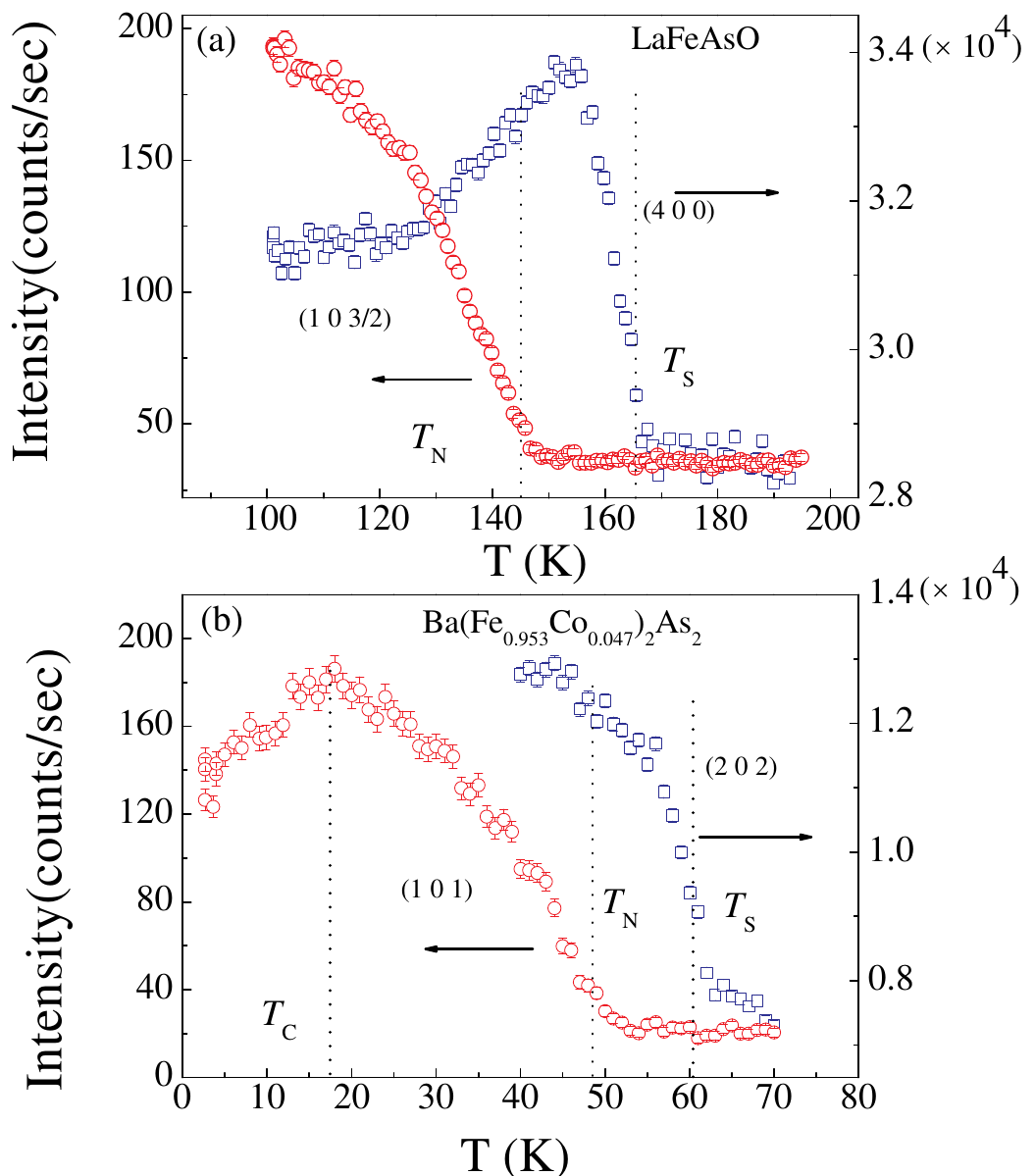} \protect\caption{(color online) 
(a) Neutron diffraction peak intensities of the (1
0 3/2) magnetic reflection and the (4 0 0)/(0 4 0) Bragg reflection
as a function of temperature in LaFeAsO. (b) Neutron diffraction peak
intensities of the (1 0 1) magnetic reflection and the (2 0 2)/(0
2 2) Bragg reflection as a function of temperature in Ba(Fe$_{0.953}$Co$_{0.047}$)$_{2}$As$_{2}$. }

\label{fig:transitions} 
\end{figure}

In LaFeAsO, neutron diffraction measurements of the (1 0 3/2) magnetic
Bragg reflection and the (4 0 0)/(0 4 0) nuclear Bragg reflection
as a function of temperature show a structural transition at $T_{{\rm S}}$=165
K split from the magnetic transition at $T_{{\rm N}}$=145 K, as illustrated
in Fig\ \ref{fig:transitions} (a), and consistent with previous
reports \cite{Yan2009,Li2011,Ramazanoglu2013,McElroy2013}. 
The (4
0 0)/(0 4 0) reflection, which develops from the (2 2 0)$_{{\rm T}}$
tetragonal Bragg reflection, was used to monitor the structural transition
indirectly by virtue of secondary extinction changes resulting from
the structural transition. 
Similarly, in Ba(Fe$_{0.953}$Co$_{0.047}$)$_{2}$As$_{2}$,
the intensity of the (2 0 2)/(0 2 2) nuclear Bragg reflection indicates
that the structural transition occurs at $T_{{\rm S}}=60$ K, which
is split from the magnetic transition at $T_{{\rm N}}=47$ K according
to the (1 0 1) magnetic Bragg reflection. The anomalous decrease of
the intensity of the (1 0 1) magnetic peak below $T_{C}\approx17$
K marks the reduction of the AFM order parameter due to competition
with the SC state \cite{Pratt2009}. The locations of these three
transitions in Ba(Fe$_{0.953}$Co$_{0.047}$)$_{2}$As$_{2}$ are
also consistent with previous reports\cite{Ni2008,Zhang2013}.

\begin{figure}
\centering \includegraphics[width=0.85\linewidth]{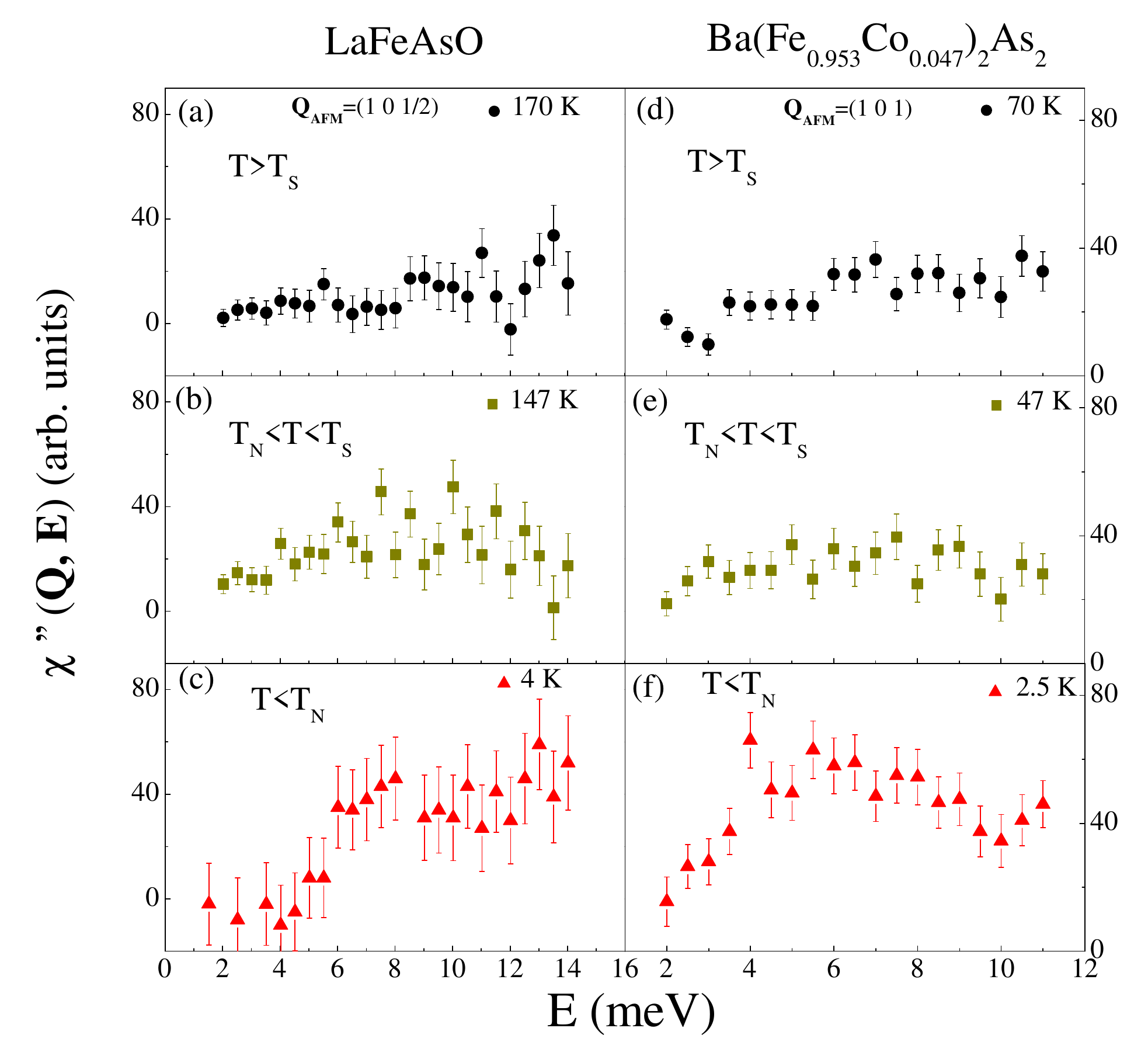} \protect\caption{(color 
online) Low-energy spin excitation in LaFeAsO at (a) 170 K,
(b) 147 K, (c) 4 K and in Ba(Fe$_{0.953}$Co$_{0.047}$)$_{2}$As$_{2}$
at (d) 70 K, (e) 47 K, (f) 2.5 K. The results are derived from the
difference between a constant-Q energy scan at \textbf{Q}$_{{\rm AFM}}$
and a background scan at \textbf{Q}$^{\prime}$=(0.7\ 0\ 1) for
Ba(Fe$_{0.953}$Co$_{0.047}$)$_{2}$As$_{2}$ and at \textbf{Q}$^{\prime}$=
(0.83\ 0\ 0.998) after the crystal was rotated from nominal \textbf{Q}$_{{\rm AFM}}$
by 20$^{{\rm o}}$ for LaFeAsO. The intensities have been normalized
to reflect a counting time of approximately five minutes. Error bars
where indicated represent one standard deviation.}

\label{fig:Escan} 
\end{figure}

To determine the impact of nematic order on the magnetic spectrum,
we explore the dependence of the imaginary part of the dynamic susceptibility
$\chi''(\textbf{Q},E)$ on the energy $E$, the momentum \textbf{Q},
and the temperature $T$. This quantity is extracted via the relationship:
\begin{equation}
S(\textbf{Q},E)\propto f^{2}(Q)\chi''(\textbf{Q},E)(1-e^{-E/k_{B}T})^{-1}\label{eq2}
\end{equation}
where $\textit{S}(\textbf{Q},E)$ is the measured background-subtracted
intensity $\textit{I}(\textbf{Q},E)-\textit{B}(\textbf{Q}^{\prime},E)$,
$\textit{f}(Q)$ is the magnetic form factor of Fe$^{2+}$, and $k_{B}$
is the Boltzmann constant. Figure\ \ref{fig:Escan} shows $\chi''(\textbf{Q}_{\textbf{AFM}},E)$
at the magnetic reflection $\textbf{Q}_{\mathrm{AFM}}=$(1\ 0\ 1/2)
in LaFeAsO and $\textbf{Q}_{\mathrm{AFM}}=$(1\ 0\ 1) in Ba(Fe$_{0.953}$Co$_{0.047}$)$_{2}$As$_{2}$
at several temperatures. Below $T_{{\rm N}}$, the spectra in LaFeAsO
exhibit the onset of an energy gap $\sim5$ meV, consistent with previous
reports\cite{Ramazanoglu2013}. In Ba(Fe$_{0.953}$Co$_{0.047}$)$_{2}$As$_{2}$,
a heavily overdamped energy gap $\sim10$ meV\cite{Tucker2014} is
observed. It has been reported\cite{Tucker2014} that upon the 
increase of Co substitution in  Ba(Fe$_{1-x}$Co$_{x}$)$_{2}$As$_{2}$, the spin 
gap appears to close gradually and is completely absent at x=0.055 due to the 
crossover from well-defined spin waves to overdamped spin excitations. The 
spin gaps in both systems vanish above $T_{{\rm N}}$
and the energy-dependent damping also increases above $T_{{\rm N}}$.
These results guide us to measure $\chi''(\mathbf{Q},E)$ at a fixed
energy transfer of $E=5$ meV in LaFeAsO and, $E=3$ meV in Ba(Fe$_{0.953}$Co$_{0.047}$)$_{2}$As$_{2}$
to obtain both the spin-spin correlation length and the magnetic intensity
as a function of temperature, according to the model for spin fluctuations
described in Ref. \cite{Tucker2014}.

\begin{figure}
\centering \includegraphics[width=0.85\linewidth]{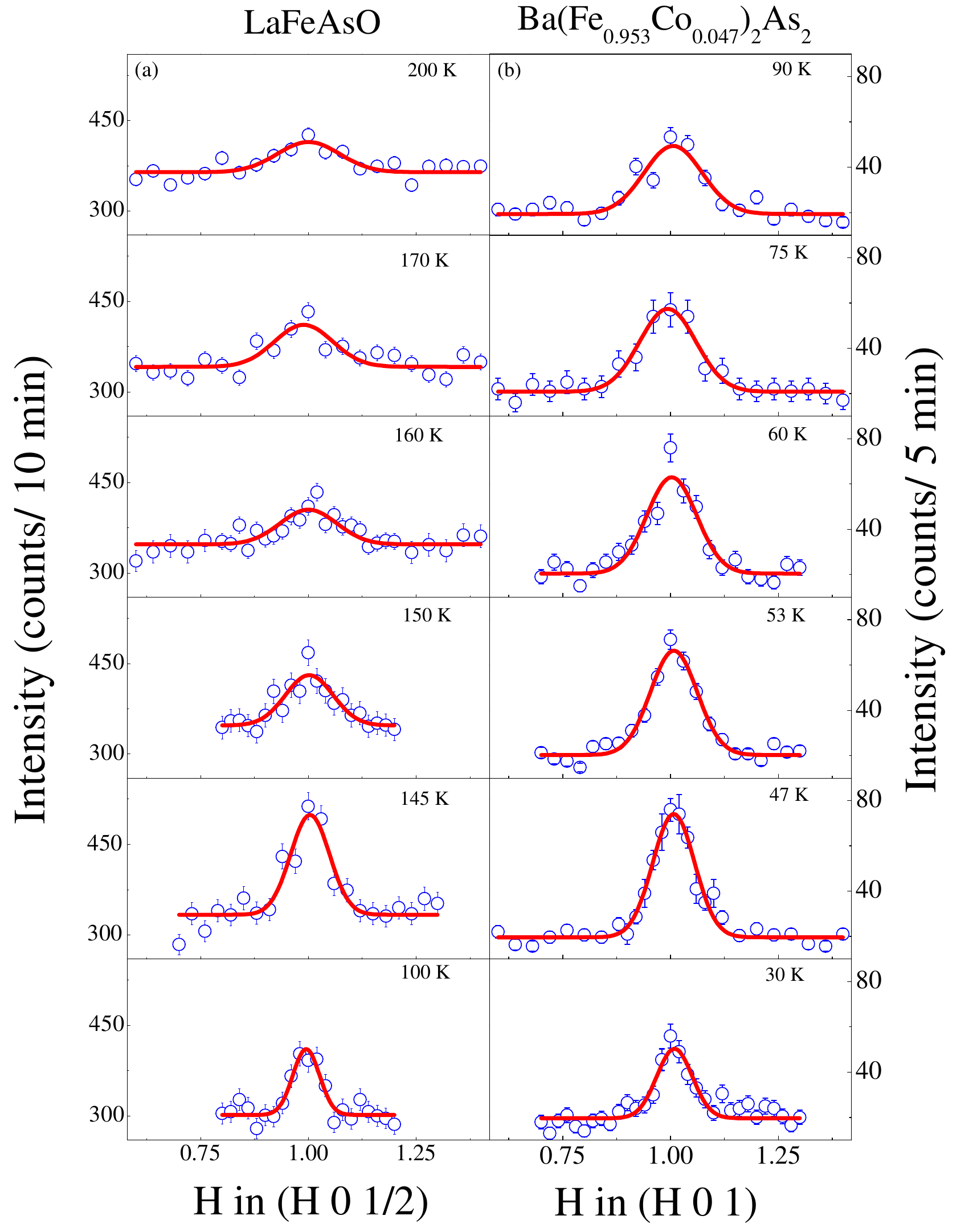} 
\protect\caption{(Color online) Representative longitudinal H scans (a) through 
\textbf{Q}$_{{\rm AFM}}$
= (1\ 0\ 1/2) at $E=5$ meV and various temperatures in LaFeAsO
and (b) through \textbf{Q}$_{{\rm AFM}}$ = (1\ 0\ 1) at $E=3$
meV and various temperatures in Ba(Fe$_{0.953}$Co$_{0.047}$)$_{2}$As$_{2}$.
The solid lines are obtained by the best fit to the data to a Gaussian
function.}

\label{fig:Hscan} 
\end{figure}

Representative longitudinal \textit{H} scans through \textbf{Q}$_{{\rm AFM}}$
= (1\ 0\ 1/2) in LaFeAsO and \textbf{Q}$_{{\rm AFM}}$ = (1\ 0\ 1)
in Ba(Fe$_{0.953}$Co$_{0.047}$)$_{2}$As$_{2}$ at low energy transfers
are shown in Fig.\ \ref{fig:Hscan}. The solid lines represent Gaussian
fits to the data,  as justified in the Supplemental Material 
\cite{SM}. We note that upon
decreasing the temperature below $T_{{\rm S}}$, the lineshape narrows
and the peak amplitude increases. The dynamic susceptibility and linewidth
(full width at half maximum) versus temperature are shown in Fig.\ \ref{fig:spincorrelation}.
Note that the reasonable mosaicity within $\sim2^{{\rm o}}$
of the coaligned LaFeAsO samples does not appreciably affect the linewidth of 
longitudinal scans and thus the linewidth reflects the intrinsic behavior of 
spin-spin correlation
length similar to that of Ba(Fe$_{0.953}$Co$_{0.047}$)$_{2}$As$_{2}$.
The dynamic susceptibility shows a discontinuous increase below $T_{{\rm S}}$
(much stronger for LaFeAsO) and exhibits a maximum at the AFM ordering
temperature $T_{{\rm N}}$, followed by a gradual decrease below $T_{{\rm N}}$
due to the opening of the spin gap. As shown in Fig.\ \ref{fig:spincorrelation}
(b) and (d), the linewidth decreases as $T$ approaches $T_{{\rm N}}$,
which is expected for a classic second-order AFM phase transition.
The striking result of this study is the observation of a sharp decrease
in the linewidth below $T_{{\rm S}}$ in both LaFeAsO and Ba(Fe$_{0.953}$Co$_{0.047}$)$_{2}$As$_{2}$
systems, which signifies a strong effect of nematic order on the approach
to AFM order.

Above the magnetic transition temperature $T_{{\rm N}}$, the linewidth
of the constant-energy Q scans
is proportional to the inverse magnetic correlation length $\xi^{-1}$
associated with the paramagnetic fluctuations \cite{Tucker2014,Kofu2009} 
(also see the Supplemental Material \cite{SM}). Therefore, the onset of 
long-range
nematic order promotes a strong increase of this correlation length,
enhancing the tendency of the system towards long-range magnetic order.
Such a cooperative interplay between nematicity and magnetism can
be understood qualitatively within models that attribute the tetragonal
symmetry-breaking to magnetic fluctuations emerging from either localized
\cite{Kivelson,Xu2008} or itinerant spins \cite{Fernandes2012b}.
To illustrate the corresponding microscopic mechanism, we show schematically
in Fig. \ref{fig:spincorrelation_theory}(a) the evolution of the
magnetic fluctuations across $T_{{\rm S}}$ and $T_{{\rm N}}$ both
in real space (upper panels) and in spin space (lower panels). The
crucial point behind this mechanism is that the iron pnictides display
two degenerate stripe AFM ground states, with ordering vectors $\mathbf{Q}_{1}=\left(1\:0\: L\right)$
and $\mathbf{Q}_{2}=\left(0\:1\: L\right)$. Thus, the magnetic ground
state can be described in terms of two interpenetrating square sublattices
-- associated with the two distinct Fe atoms in the unit cell -- that
tend to order magnetically in Néel-like configurations (blue and red
dashed lines in Fig. \ref{fig:spincorrelation_theory}(a)).

\begin{figure}
\centering \includegraphics[width=0.85\linewidth]{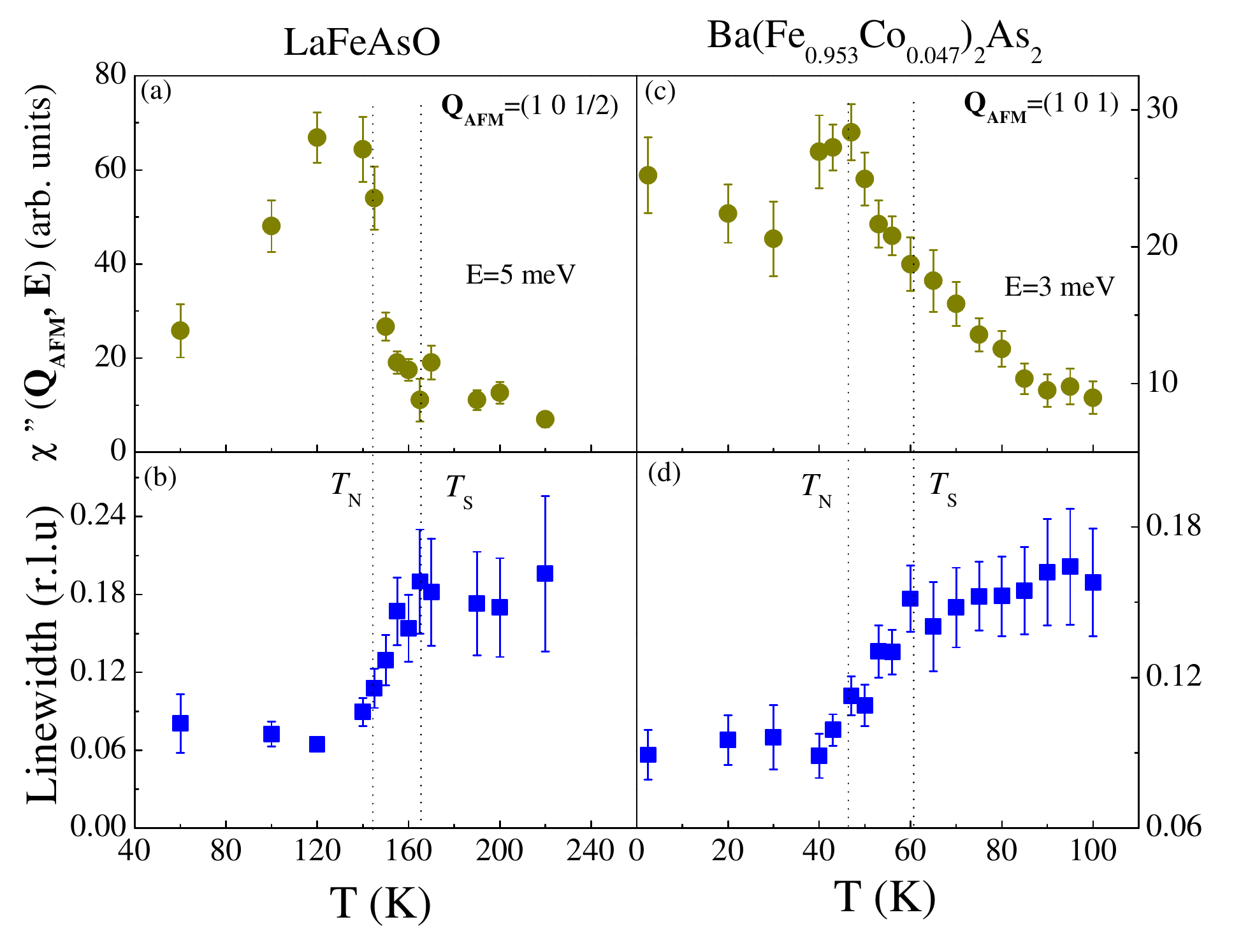} 
\protect\caption{(color online) Temperature dependence of (a) dynamic 
susceptibility
$\chi''(\textbf{Q},E=5$ meV) and (b) the Gaussian linewidth, obtained
by fitting the longitudinal H scans through \textbf{Q}$_{{\rm AFM}}$
= (1\ 0\ 1/2) at $E=5$ meV in LaFeAsO. Temperature dependence of
(c) the dynamic susceptibility $\chi''(\textbf{Q},E=3$ meV) and (d)
the Gaussian linewidth, obtained by fitting the longitudinal H scans
through \textbf{Q}$_{{\rm AFM}}$ = (1\ 0\ 1) at $E=3$ meV in Ba(Fe$_{0.953}$Co$_{0.047}$)$_{2}$As$_{2}$.
The vertical dashed lines mark the locations of the structural transition
$T_{{\rm S}}$ and the AFM magnetic transition $T_{{\rm N}}$.}

\label{fig:spincorrelation} 
\end{figure}

Above $T_{{\rm S}}$, where there is no long-range magnetic order,
these two sublattices are essentially independent (as shown in the
upper left panel of Fig. \ref{fig:spincorrelation_theory}(a)), and
their fluctuations are uncoupled (as shown in the lower left panel).
As a result, the system has multiple possible ground states, a feature
commonly seen in frustrated spin systems with low magnetic transition
temperatures. However, below $T_{{\rm S}}$ but above $T_{{\rm N}}$,
nematic order emerges as a coupling between the two sublattices (upper
middle panel in Fig.\ref{fig:spincorrelation_theory}(a)), enforcing
the two corresponding Néel order parameters to fluctuate coherently
either anti-parallel (as shown in the lower middle panel) or parallel
to each other. There is still no long-range magnetic order, since
the spins can point at any direction in spin space. However, the tetragonal
symmetry of the system is broken, since nearest-neighbor spins are
locked in a ferromagnetic-like or an antiferromagnetic-like configuration.
Furthermore, by breaking the tetragonal symmetry, nematic order reduces
the number of possible magnetic ground states to only one -- either
the $\mathbf{Q}_{1}=\left(1\:0\: L\right)$ stripe if the $a$ direction
is selected along the $x$ axis, or the $\mathbf{Q}_{2}=\left(0\:1\: L\right)$
stripe if the $a$ direction is selected along the $y$ direction.
Thus, the frustration, resulting from two degenerate magnetic stripe states 
present at higher temperatures, is lifted by
nematic order, leading to an enhancement of the spin-spin correlation
length $\xi$, and therefore of $T_{{\rm N}}$, which sets in when
$\xi$ diverges (right panels). Note that this phenomenon can be observed
even in twinned samples as the ones studied here, since magnetic fluctuations
are enhanced regardless of the type of nematic domain selected.

To go beyond this qualitative analysis, we calculate
$\xi$ using a low-energy action for the magnetic degrees of freedom
that accounts for the existence of two symmetry-related magnetic instabilities
which give rise to a preemptive nematic phase at $T_{S}>T_{N}$ (see
Ref. \cite{Fernandes2012b} for a microscopic derivation from an intinerant
3-band model). The equations for $\xi$ and the parameters used here
are presented in the Supplemental Material \cite{SM}. To take into account
the resolution limitations in the linewidth $W$ imposed by the instrument
and by the fact that the measurements are performed at non-zero energy,
we shift $\xi^{-1}$ by a temperature-independent term $\delta_{\mathrm{res}}>0$,
$W\propto\xi^{-1}+\delta_{\mathrm{res}}$. The results are shown in
Fig. \ref{fig:spincorrelation_theory}(b). Because our model is based
on an expansion near $T_{N}$, it systematically underestimates the
correlation length at higher temperatures. Yet, it captures the main
qualitative feature observed experimentally, namely, the sharp enhancement
of $\xi$ below $T_{S}$ due to the onset of long-range nematic order.
This is shown explicitly in Fig. \ref{fig:spincorrelation_theory}b
by comparing the hypothetical behavior of $\xi$ in the absence of
nematic order (dashed lines) with the behavior in the presence of
nematicity (solid lines). We emphasize that this theoretical calculation
is intended to highlight the strong feedback effect of nematic order
on the magnetic fluctuations, and not to capture the full quantitative
dependence of $\xi$ on temperature, which will be affected by other
features such as domains, mosaicity, etc. 

\begin{figure}
\centering \includegraphics[width=0.8\linewidth]{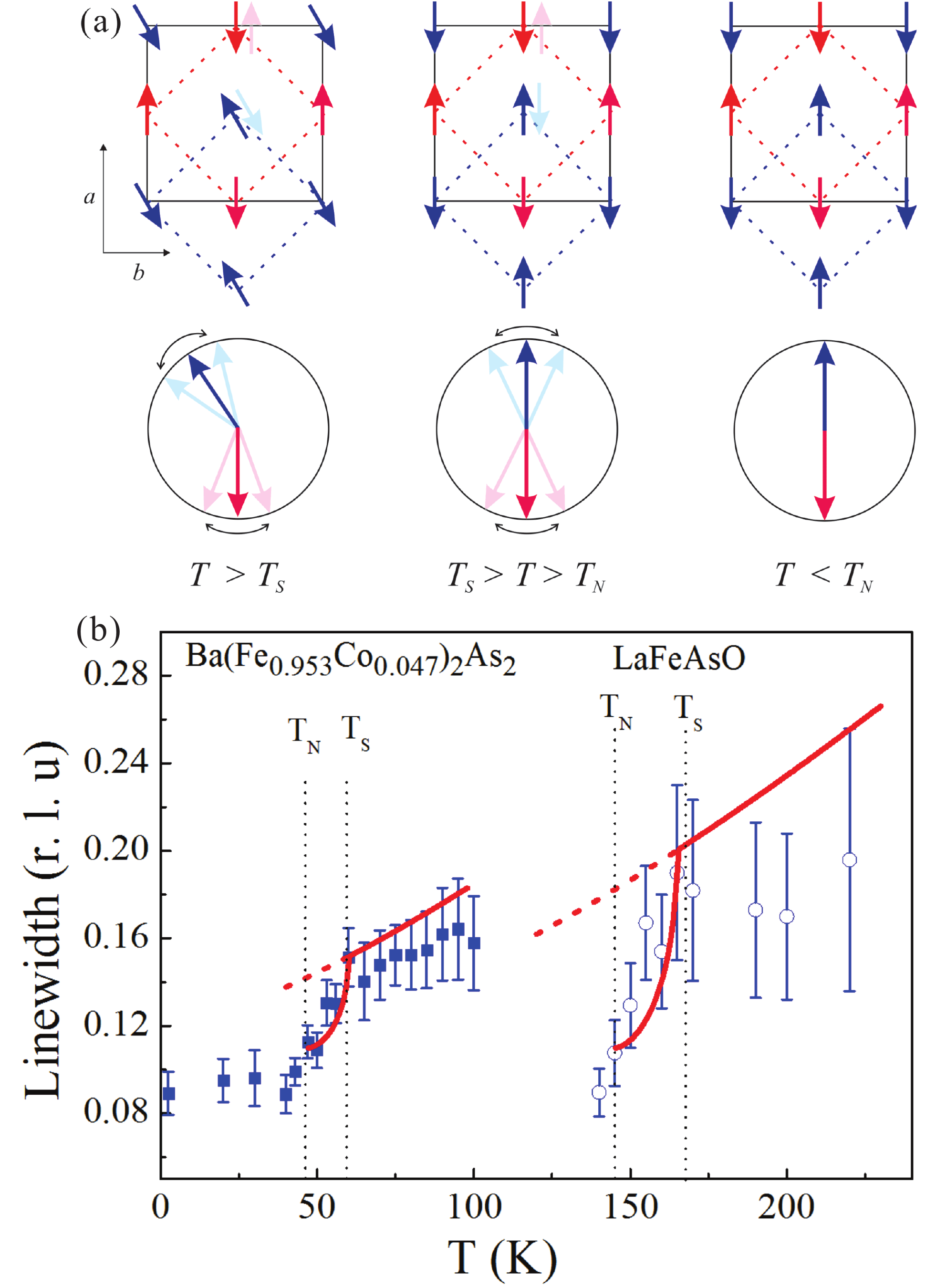} \protect\caption{(color 
online) (a) Evolution of the magnetic fluctuations of the iron
pnictides in real space (upper panels) and spin space (lower panels).
The two Néel sublattices corresponding to the two different Fe atoms
of the unit cell (dashed lines) are shown in red and blue. Above $T_{{\rm N}}$,
spins in each sublattice fluctuate around a Néel configuration. These
fluctuations are uncoupled above $T_{{\rm S}}$, but below $T_{{\rm S}}$,
the two fluctuating Néel sublattices are coupled either parallel or
anti-parallel to each other. The double arrows in the upper panels
represent fluctuating spins, as explicitly shown in the lower panels.
Below $T_{{\rm N}}$, spins point to a fixed direction in spin space.
(b) Temperature dependence of the theoretical linewidth $W$ (red
lines) compared to the experimental linewidth (blue dots). The dashed
lines mark the locations of $T_{{\rm S}}$ and $T_{{\rm N}}$ in LaFeAsO
and Ba(Fe$_{0.953}$Co$_{0.047}$)$_{2}$As$_{2}$. }

\label{fig:spincorrelation_theory} 
\end{figure}

In summary, we have reported unambiguous evidence for the feedback
effect of nematic order on the magnetic spectrum in both ``1111''
and underdoped ``122'' families of the iron pnictides with $T_{S}>T_{N}$,
manifested by the sharp enhancement of the spin-spin correlation length
below $T_{{\rm S}}$, revealing a key impact of this elusive electronic
order on the normal-state properties of the iron arsenides. Since
magnetic fluctuations are believed to be important for the formation
of the SC state \cite{reviews_pairing}, and our results provide evidence
that nematic order enhances them, this suggests that nematicity may
be more than another competing order, as previously reported \cite{Nandi10,Bohmer12},
and may even help enhancing $T_{{\rm C}}$ in some circumstances \cite{Fernandes_Millis13,DungHaiLee13}.

\emph{Acknowledgements.} Research at Ames Laboratory is supported
by the US Department of Energy, Office of Basic Energy Sciences, Division
of Materials Sciences and Engineering under Contract No. DE-AC02-07CH11358.
R.M.F. is supported by the Department of Energy under Award Number
DE-SC0012336. Use of the high flux isotope reactor at the Oak Ridge
National Laboratory, was supported by the US Department of Energy,
Office of Basic Energy Sciences, Scientific User Facilities Division.
The NIST Center for Neutron Research is supported by the US Department
of Commerce. We acknowledge Dan Parshall for his technical assistance
in measuring Ba(Fe$_{0.953}$Co$_{0.047}$)$_{2}$As$_{2}$ at BT-7
triple-axis neutron spectrometer at the NIST center for Neutron Research.

\clearpage

\textbf{Supplementary material for: ``Sharp enhancement of spin fluctuations
by nematic order in iron pnictides''}\begin{LARGE}                              
 
   \end{LARGE}
%
%

\author{Qiang Zhang}

\affiliation{Ames Laboratory, Ames, IA, 50011, USA}

\affiliation{Department of Physics and Astronomy, Iowa State University, Ames,
IA, 50011, USA}

\author{Rafael M. Fernandes}

\affiliation{School of Physics and Astronomy, University of Minnesota, 
Minneapolis,
MN 55455, USA}

\author{Jagat Lamsal}

\affiliation{Ames Laboratory, Ames, IA, 50011, USA}

\affiliation{Department of Physics and Astronomy, Iowa State University, Ames,
IA, 50011, USA}

\author{Jiaqiang Yan}

\affiliation{Oak Ridge National Laboratory, Oak Ridge, Tennessee 37831, USA}

\author{Songxue Chi}

\affiliation{Oak Ridge National Laboratory, Oak Ridge, Tennessee 37831, USA}

\author{Gregory S. Tucker}

\affiliation{Ames Laboratory, Ames, IA, 50011, USA}

\affiliation{Department of Physics and Astronomy, Iowa State University, Ames,
IA, 50011, USA}

\author{ Daniel K. Pratt}

\affiliation{NIST Center for Neutron Research, National Institute of Standards
and Technology, Gaithersburg, Maryland 20899-6102, USA}

\author{Jeffrey W. Lynn}

\affiliation{NIST Center for Neutron Research, National Institute of Standards
and Technology, Gaithersburg, Maryland 20899-6102, USA}

\author{R. W. McCallum}

\affiliation{Ames Laboratory, Ames, IA, 50011, USA}

\affiliation{Department of Materials Sciences and Engineering, Iowa State 
University,
Ames, Iowa 50011, USA}

\author{Paul C. Canfield}

\affiliation{Ames Laboratory, Ames, IA, 50011, USA}

\affiliation{Department of Physics and Astronomy, Iowa State University, Ames,
IA, 50011, USA}

\author{Thomas A. Lograsso}

\affiliation{Ames Laboratory, Ames, IA, 50011, USA}

\affiliation{Department of Materials Sciences and Engineering, Iowa State 
University,
Ames, Iowa 50011, USA}

\author{Alan Goldman}

\affiliation{Ames Laboratory, Ames, IA, 50011, USA}

\affiliation{Department of Physics and Astronomy, Iowa State University, Ames,
IA, 50011, USA}

\author{David Vaknin}

\affiliation{Ames Laboratory, Ames, IA, 50011, USA}

\affiliation{Department of Physics and Astronomy, Iowa State University, Ames,
IA, 50011, USA}

\author{Robert J. McQueeney}

\affiliation{Ames Laboratory, Ames, IA, 50011, USA}

\affiliation{Department of Physics and Astronomy, Iowa State University, Ames,
IA, 50011, USA}

\affiliation{Oak Ridge National Laboratory, Oak Ridge, Tennessee 37831, USA}

\date{\today}

\maketitle

\section{Linewidth fits}

In this Supplemental section, we provide more details on the justification
of the Gaussian fit performed to the constant-energy longitudinal
scans presented in Fig. 3 of the article. In the main text, we identified
the Gaussian linewidth with the inverse magnetic correlation length.
This is justified because the energy probed is much smaller than the
damping of the magnetic excitations present in the paramagnetic state.

To make this point clearer, we use the microscopically-derived diffusive
model discussed in Refs. \cite{Diallo10,Li10,Tucker12,Tucker14} that
captures the low-energy magnetic excitations near the magnetic ordering
vector $\mathbf{Q}$. Within this model, the in-plane dynamic magnetic
susceptibility is given by (in tetragonal notation):

\begin{equation}
\chi\left(\mathbf{q}+\mathbf{Q_{AFM}},E\right)=\frac{\chi_{0}}{a^{2}\left(\xi^{
-2}+q_{x}^{2}+q_{y}^{2}+\eta q_{x}q_{y}\right)-iE\gamma}\label{eq1}
\end{equation}
where $\chi_{0}^{-1}$ is an overall magnetic energy scale, $\xi$
is the magnetic correlation length, $E$ is the energy, $\mathbf{q}$
is the reduced momentum, $a$ is the lattice constant, $\eta$ is
the in-plane anisotropy parameter, and $\gamma$ is the Landau damping.
This expression is derived from an effective three-band model and
ultimately relies on the fact that the paramagnetic excitations can
decay into particle-hole excitations, giving rise to Landau damping.
The comparison with the spin-spin correlation function 
$S\left(\mathbf{Q},E\right)$,
extracted in the inelastic neutron scattering (INS) experiments, is
achieved via the fluctuation-dissipation theorem:

\begin{equation}
S\left(\mathbf{Q},E\right)\propto\left(1-\mathrm{e}^{-E/k_{B}T}\right)^{-1}
\mathrm{Im}\left[\chi\left(\mathbf{Q},E\right)\right]\label{eq2}
\end{equation}

\begin{figure}
\centering \includegraphics[width=0.7\linewidth]{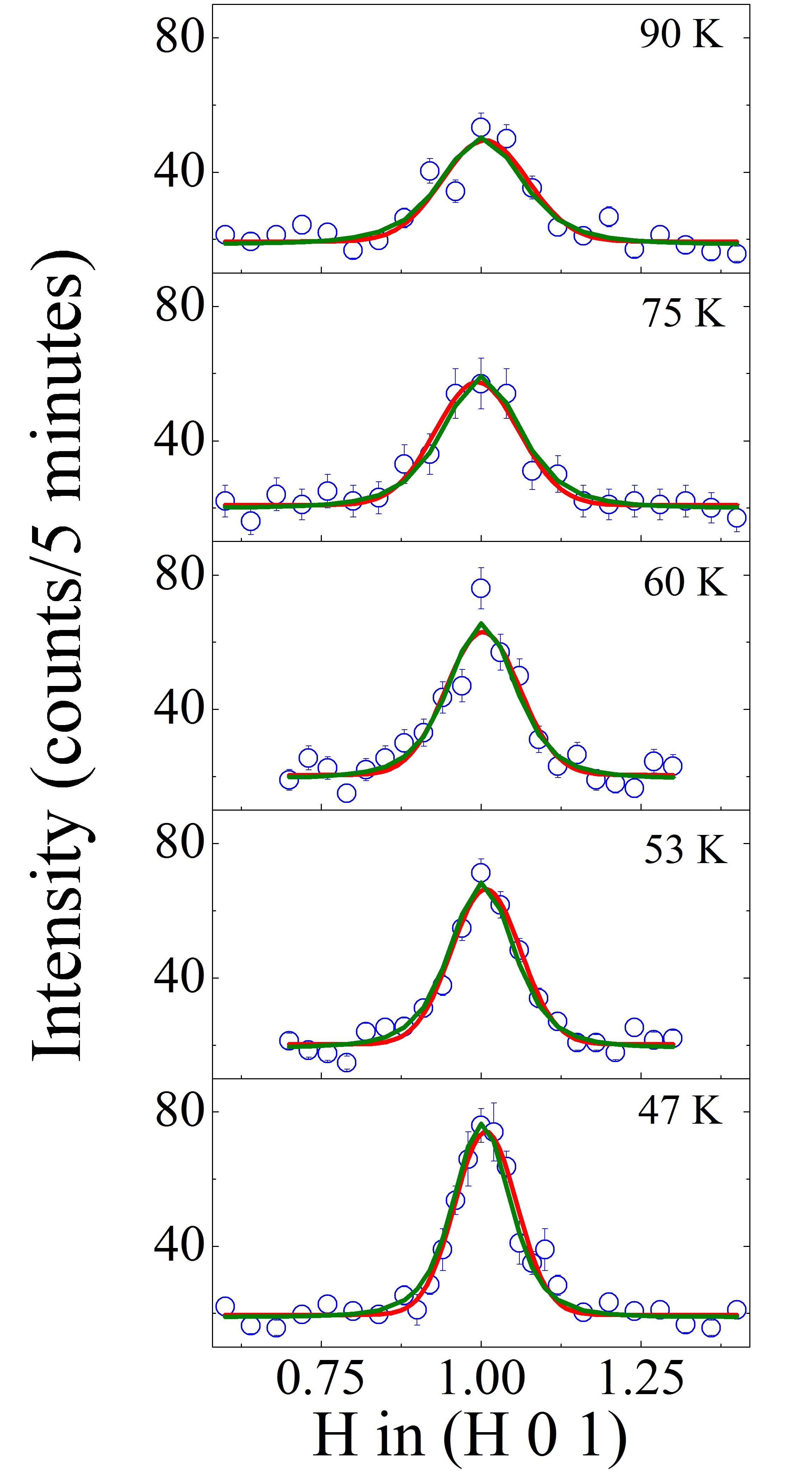}
\protect\caption{(color online) Comparison of two types of fits using a 
Gaussian 
function
(red solid line) and the diffusive model of Eq. (\ref{eq1}) (olive
solid line) to the experimental data (open circles) in 
Ba(Fe$_{0.953}$Co$_{0.047}$)$_{2}$As$_{2}$. }

\label{fig:twofits} 
\end{figure}

Previously, Eq.(\ref{eq1}) has been successfully employed to fit
the experimental INS data in the paramagnetic state across the entire
phase diagram of the $\mathrm{Ba\left(Fe_{1-x}Co_{x}\right)_{2}As_{2}}$
compounds \cite{Li10,Tucker12,Tucker14}. In particular, the only
temperature-dependent parameter is the correlation length $\xi$,
while $\eta$ and $\gamma$ depend only on the Co concentration $x$.

\begin{figure}
\centering \includegraphics[width=0.95\linewidth]{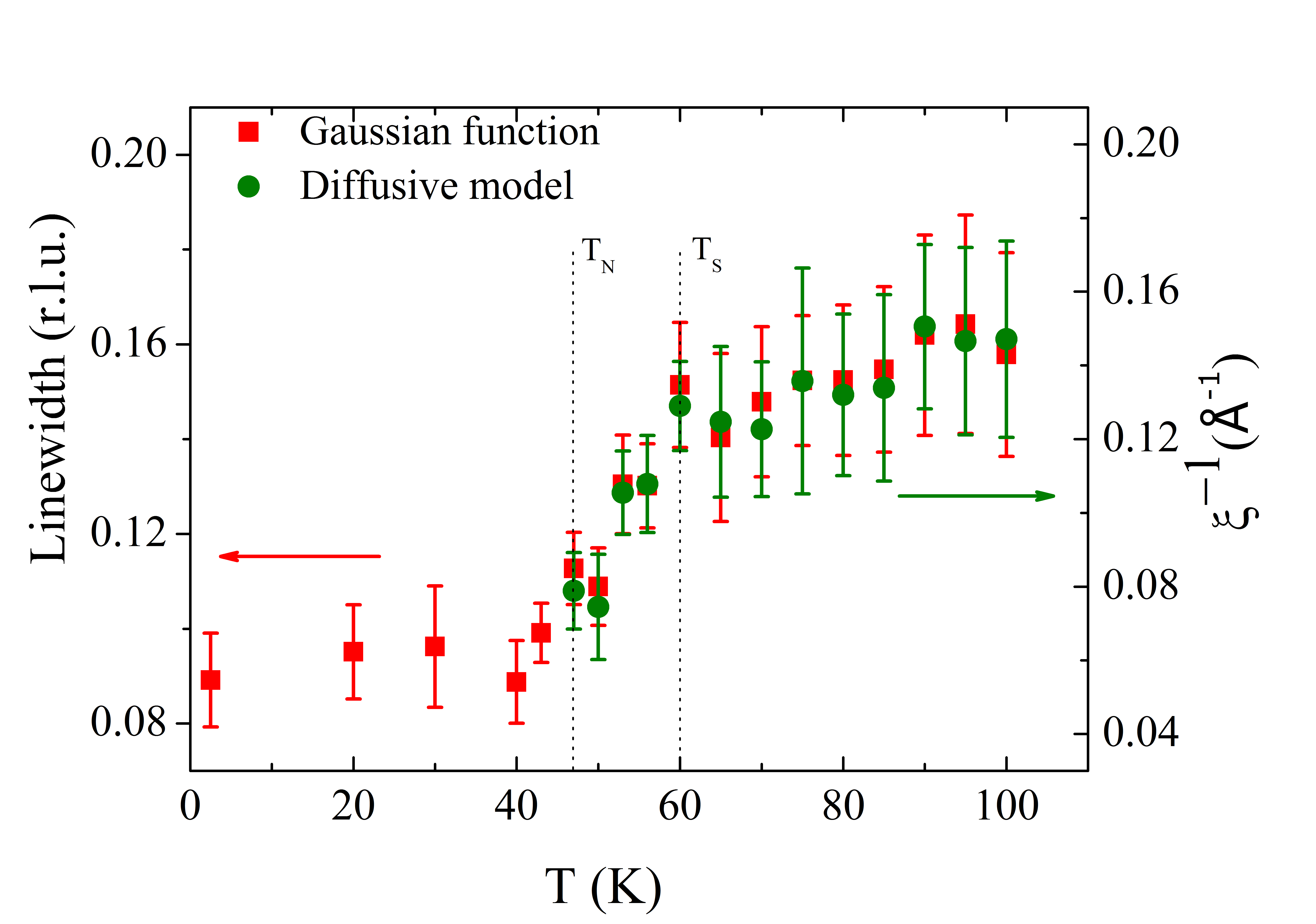}
\protect\caption{(color online) Comparison of the temperature dependence of the 
inverse
spin-spin correlation length $\xi^{-1}$ derived from the diffusive
model of Eq. (\ref{eq1}) (olive circles) and the linewidth obtained
from the Gaussian function fits (red squares) in 
Ba(Fe$_{0.953}$Co$_{0.047}$)$_{2}$As$_{2}$.
The vertical dashed lines mark the positions of $T_{{\rm S}}$ and
$T_{{\rm N}}$.}

\label{fig:comparisonxi} 
\end{figure}

To check whether the effective Gaussian model used to fit the data
of the $x=0.047$ sample (taken at $E=3$ meV and shown in the right
column of Fig. 3 of the main text) is consistent with the 
microscopically-derived
diffusive model, we use the temperature-independent parameters reported
in Ref. \cite{Tucker12} for $x=0.047$, $\eta=1.14$ and $\gamma^{-1}=75$
meV, and extract the temperature dependence of the magnetic correlation
length $\xi$ by fitting the experimental INS intensity corrected
for the Bose thermal population factor and the Fe$^{2+}$ single-ion
magnetic form factor to Eq. (\ref{eq1}) after convolution with the
Popovici approximation to the instrumental resolution using the RESLIB
program \cite{Zheludev}. The fits for several temperatures are shown
in Fig.\ \ref{fig:twofits}, and the temperature dependence of the
inverse spin-spin correlation length is shown in Fig.\ \ref{fig:comparisonxi}
. Comparison with the fits to the effective Gaussian model are also
presented in both figures, revealing that indeed the Gaussian linewidth
correctly captures the temperature dependence of the inverse correlation
length and, in particular, its sharp suppression below $T_{{\rm S}}$.
Note that because this model is appropriate only for the paramagnetic
phase, the fittings were only performed above $T_{{\rm N}}$.

The reason behind this agreement between the two models can be understood
directly from Eqs. (\ref{eq1}) and (\ref{eq2}). In particular, since
$E\gamma\approx0.04\ll1$, the behavior of 
$\mathrm{Im}\left[\chi\left(\mathbf{q},E\right)\right]$
is dominated by the static part, which, due to the convolution with
the experimental resolution, is well captured by an effective Gaussian
curve. For $\mathrm{LaFeAsO}$, a systematic fitting of the INS data
in the paramagnetic state to the diffusive model Eq. (\ref{eq1})
is not available. Nevertheless, because the damping factors in the
magnetically ordered states of both $\mathrm{LaFeAsO}$ and 
$\mathrm{Ba\left(Fe_{0.953}Co_{0.047}\right)_{2}As_{2}}$
have similar magnitudes \cite{Ramazanoglu13}, it is not unreasonable
to expect the same to be true in the paramagnetic state. In this case,
$E\gamma\ll1$ would also be true in $\mathrm{LaFeAsO}$, justifying
the use of an effective Gaussian curve to fit the constant-energy
longitudinal scans.

\section{Theoretical model}

The theoretical model presented in the main text for the 
temperature
dependence of the correlation length $\xi$ was derived previously
in Ref. \cite{Fernandes2012b}. The key ingredient of this model is
the existence of two magnetic instabilities at the ordering vectors
$\mathbf{Q}_{1}=\left(\pi,0\right)$ and $\mathbf{Q}_{2}=\left(0,\pi\right)$
(in units of the square Fe lattice parameter). The two corresponding
order parameters are denoted by $\mathbf{M}_{1}$ and $\mathbf{M}_{2}$,
and the magnetic action is given by:

\begin{eqnarray}
S\left[\mathbf{M}_{i}\right] & = & 
\int_{q}\chi_{q}^{-1}\left(M_{1}^{2}+M_{2}^{2}\right)+\frac{u}{2}\int_{x}
\left(M_{1}^{2}+M_{2}^{2}\right)^{2}\nonumber \\
 &  & -\frac{g}{2}\int_{x}\left(M_{1}^{2}-M_{2}^{2}\right)^{2}\label{eq_S}
\end{eqnarray}
where $\int_{q}=T\sum_{\omega_{n}}\int\frac{d^{d}q}{\left(2\pi\right)^{d}}$
and $\int_{x}=\int_{0}^{\beta}d\tau\int d^{d}x$. Here $u>g>0$ are
phenomenological parameters that can in principle be derived from
a microscopic 3-band model \cite{Fernandes2012b}, and 
$\chi_{q}^{-1}=r_{0}+q^{2}$
for a classical phase transition, with $r_{0}$ a temperature-dependent
tuning parameter. Within this model, the nematic order parameter,
given by $\varphi=g\left\langle M_{1}^{2}-M_{2}^{2}\right\rangle $,
can condense at a temperature above the magnetic transition temperature,
breaking the tetragonal symmetry of the system, since $\mathbf{M}_{1}$
and $\mathbf{M}_{2}$ are related by a $90^{\circ}$ rotation. Thus,
from this action, one can derive the behavior of the magnetic correlation
length $\xi$ across the nematic phase transition. In the large-$N$
approach, where Gaussian magnetic fluctuations are included self-consistently,
one obtain two coupled non-linear equations for the parameters $r$
and $\varphi$:

\begin{eqnarray}
r & = & 
\bar{r}_{0}-\frac{\bar{u}}{4}\left[\left(r+\varphi\right)^{\frac{d-2}{2}}
+\left(r-\varphi\right)^{\frac{d-2}{2}}\right]\nonumber \\
\varphi & = & 
\frac{\bar{g}}{4}\left[\left(r+\varphi\right)^{\frac{d-2}{2}}
-\left(r-\varphi\right)^{\frac{d-2}{2}}\right]\label{eq_self_cons}
\end{eqnarray}
where $d$ is the dimensionality and $\bar{u}$, $\bar{g}$, $\bar{r}_{0}$
are the corresponding renormalized parameters of the original action.
The magnetic correlation length can be obtained via $\xi^{-2}\propto r-\varphi$.
It is clear, in this regard, the origin of the kink observed in $\xi$:
it arises because, above $T_{s}$, $\varphi=0$, whereas below $T_{S}$,
$\varphi\neq0$. In particular, introducing the auxiliary variable
$z=\varphi/r$, the magnetic correlation length is given by:

\begin{equation}
\xi^{-1}=A\sqrt{\left[\frac{\left(1+z\right)^{\frac{d-2}{2}}-\left(1-z\right)^{
\frac{d-2}{2}}}{z}\right]^{\frac{2}{4-d}}\left(1-z\right)}\label{eq_xi}
\end{equation}
where $A$ is a positive constant and $z$ is determined implicitly
as function of 
$\bar{r}_{0}\rightarrow\bar{r}_{0}/\left(\frac{g}{4}\right)^{\frac{2}{4-d}}$
according to:

\begin{eqnarray}
\bar{r}_{0} & = & 
\left[\frac{\left(1+z\right)^{\frac{d-2}{2}}-\left(1-z\right)^{\frac{d-2}{2}}}{z
}\right]^{\frac{d-2}{4-d}}\times\label{eq_z}\\
 &  & 
\left[\left(1+z\right)^{\frac{d-2}{2}}\left(\alpha+\frac{1}{z}
\right)+\left(1-z\right)^{\frac{d-2}{2}}\left(\alpha-\frac{1}{z}\right)\right]
\nonumber 
\end{eqnarray}
where $\alpha\equiv u/g$. As shown explicitly in Ref. \cite{Fernandes2012b},
to mimic the interlayer coupling in the iron pnictides, one can consider
an intermediate dimensionality $2<d<3$. The theoretical results presented
in Fig. 5 of the main text were obtained by solving Eqs. (\ref{eq_xi})
and (\ref{eq_z}) for $d=2.6$ and the following set of parameters:
for LaFeAsO, we used $\alpha=17$, $\bar{r}_{0}=0.18\left(T-9.3\right)$,
and $A=0.13$; for Ba(Fe$_{0.953}$Co$_{0.047}$)$_{2}$As$_{2}$
we used $\alpha=30$, $\bar{r}_{0}=0.54\left(T+30.9\right)$, and
$A=0.06$.

As explained in the main text and in the previous section of the Supplemental
Material, the linewidth $W$ measured experimentally is limited by
both the instrument resolution and by the fact that the measurements
were performed at non-zero energies $E>0$. For instance, from Eq.
(\ref{eq1}) we note that even when $\xi\rightarrow\infty$, the spin-spin
correlation function acquires effectively a finite linewidth, according
to:

\begin{equation}
\frac{\chi''\left(\mathbf{q}+\mathbf{Q},E\right)}{\omega}=\frac{\chi_{0}\gamma}{
\left(a^{2}q^{2}\right)^{2}+\gamma^{2}E^{2}}\label{W_res}
\end{equation}

For these reasons, the measured linewidth does not become zero at
the magnetic transition, but instead becomes a constant $\delta_{\mathrm{res}}$.
To capture this effect in a simple way, we considered a uniform shift
of the linewidth, $W=\xi^{-1}+\delta_{\mathrm{res}}$, with 
$\delta_{\mathrm{res}}=0.11$
in both cases.

\end{document}